\documentclass{article}

\usepackage{icrctc07}

\newcommand{\Xmax}{\ensuremath X_\mathrm{max}}
\newcommand{\meanXmax}{\ensuremath \langle X_\mathrm{max}\rangle}
\newcommand{\gcm}{g\,cm$^{-2}$}
\def\EeV{\ifmmode {\mathrm{\ Ee\kern -0.1em V}}\else
                   \textrm{Ee\kern -0.1em V}\fi}%
\def\eV{\ifmmode {\mathrm{\ e\kern -0.1em V}}\else
                   \textrm{e\kern -0.1em V}\fi}%

\title{Study of the Cosmic Ray Composition above 0.4 \EeV{} using the
Longitudinal Profiles of Showers observed at the Pierre Auger
Observatory}

\shorttitle{Study of the Cosmic Ray Composition with the Pierre Auger Observatory}

\authors{Michael Unger $^{1}$ for the Pierre Auger Collaboration$^{2}$}
\shortauthors{Michael Unger [Pierre Auger Collaboration]}
\afiliations{$^1$Forschungszentrum Karlsruhe, Postfach 3640, 76021 Karlsruhe, Germany\\ 
             $^2$Observatorio Pierre Auger, Av. San Martin Norte 304, 5613 Malarg\"ue, Argentinia}
\email{Michael.Unger@ik.fzk.de}

\abstract{The Pierre Auger Observatory has been collecting data in a stable manner since January 2004. We present
here a study of the cosmic ray composition using events recorded in hybrid mode during the first years
of data taking. These are air showers observed by the fluorescence detector as well as the surface detector, so
the depth of shower maximum, $\Xmax$, is measured directly. The cosmic ray composition is studied in
different energy ranges by comparing the observed average $\Xmax$ with predictions from air shower
simulations for different nuclei. The change of $\meanXmax$ with energy (elongation rate) is used to derive
estimates of the change in primary composition.}
\begin{document}
\maketitle
\section{Introduction}
Ultra-high-energy cosmic rays are presumed to be of extragalactic origin. 
With increasing energies, and thus Larmor radii, the galactic charged particles can not be 
confined in our Galaxy and the galactic cosmic ray accelerator candidates are expected to 
reach their maximum energy well below 10$^{18}$~\eV.  Moreover, there are no experimental
signs of an anisotropy of the cosmic ray arrival direction at these energies.\\
The transition between galactic and extragalactic cosmic rays is therefore 
believed to happen between 10$^{18}$ and 10$^{19}$ \eV{}
where a spectral break in the cosmic ray flux known as the 'ankle' or 'dip'
is observed. The exact position and nature of the transition is still disputed and it seems
clear that a combined precise measurement of the particle flux and composition in this
energy range is needed to be able to distinguish between different models of the
extragalactic cosmic ray component (see [1] for recent discussions on this subject).\\
For fluorescence detectors (FDs), the observable most sensitive to the
composition is the slant depth position $\Xmax$ at which the maximum of the longitudinal shower profile 
occurs. Its average value $\meanXmax$ at a certain energy $E$ is related to the mean logarithmic mass $\langle\ln A\rangle$
via 
\begin{equation}
   \meanXmax = D_\mathrm{p}\left[\ln\left( E/E_0\right)-\langle\ln A\rangle\right] + \mathrm{c}_p,
 \label{eq:elong}
\end{equation}
where $D_\mathrm{p}$ denotes the 'elongation rate' [2] of a proton, and $\mathrm{c}_p$ is the average
depth of a proton with reference energy $E_0$. Both, $D_\mathrm{p}$ and $\mathrm{c}_p$, depend
 on the nature of hadronic interactions. The width of the $\Xmax$ distribution is another composition
sensitive parameter, since heavy nuclei are expected to produce smaller shower-to-shower fluctuations than protons. 

\section{Data Analysis}
\begin{figure}
  \includegraphics[clip,width=.5\textwidth]{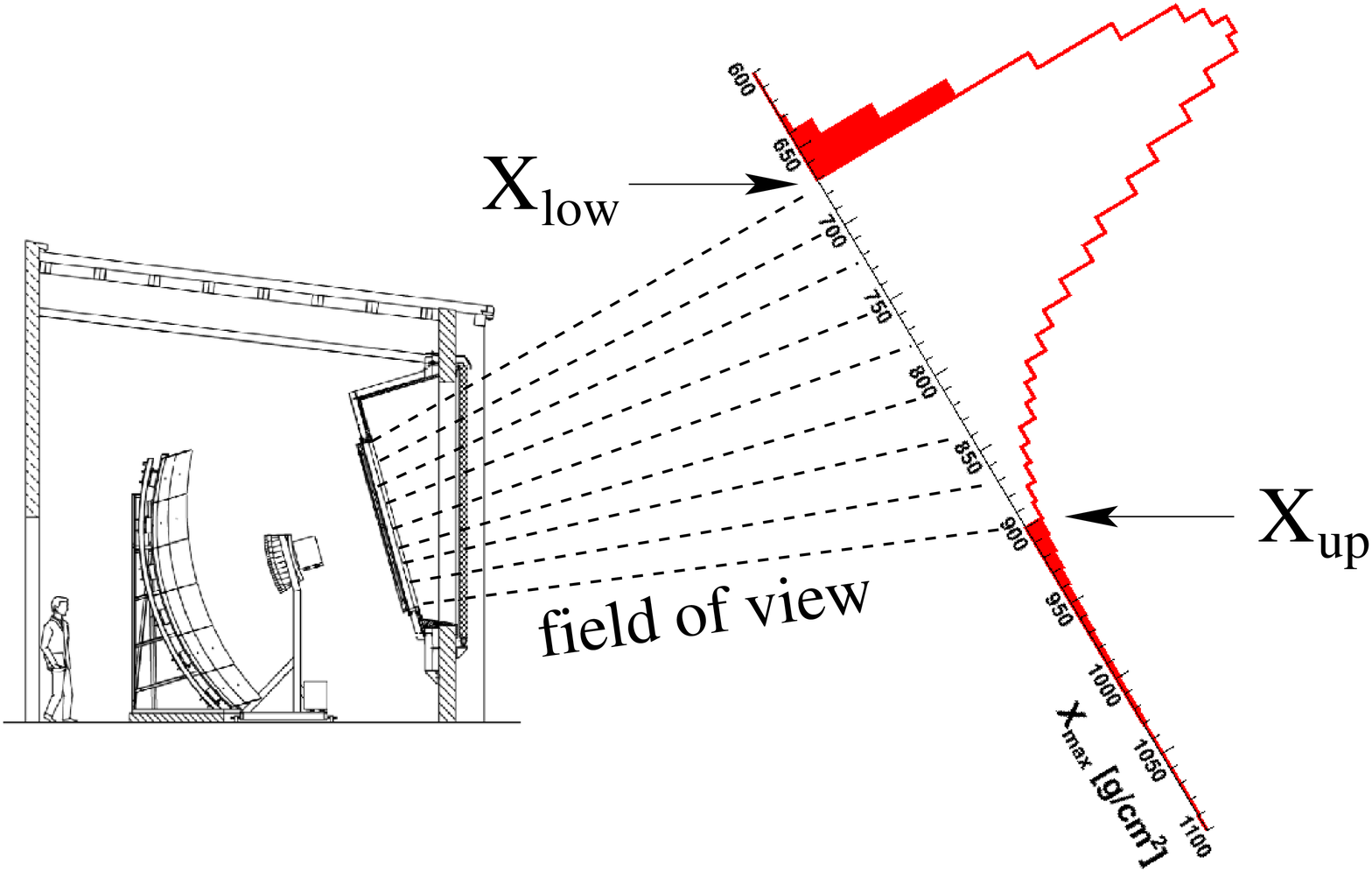}
\caption{Illustration of the effect of the field of view of the fluorescence detector
                             on the selected $\Xmax$ distribution. Filled areas indicate 
         slant depths, which are de-selected by the quality cuts.}
  \label{fig:fov1}
\end{figure}
In this analysis we use hybrid events collected by the Pierre Auger Observatory between the 1st of December 2004 and 
the 30th of April 2007. 
These are showers observed by at least one FD and with at least one triggered tank recorded by
the surface detector.\\
In order to ensure a good $\Xmax$ resolution at the 20~\gcm{} level [3], the following
quality cuts were applied to the data: The reconstructed $\Xmax$ should lie within the observed
shower profile and the reduced $\chi^2$ of a fit with a Gaisser-Hillas function~[4] should not exceed 2.5. Moreover,
insignificant shower maxima are rejected by requiring that the $\chi^2$ of 
a linear fit to the longitudinal profile exceeds the Gaisser-Hillas fit $\chi^2$ by at least four. Finally,
the estimated uncertainties of the shower maximum and total energy must be smaller than 40~\gcm{} and 20\%, respectively.

\begin{figure}
  \includegraphics[clip,width=.5\textwidth]{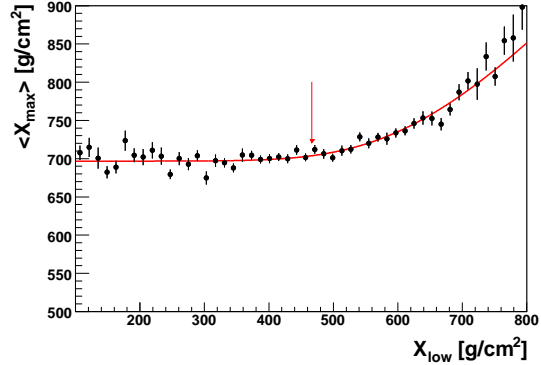}
  \caption{Dependence of the average measured $\Xmax$ on the upper viewable slant depth boundary for showers with 
energies between 10$^{18}$ and 10$^{18.25}$ \eV{}. The arrow indicates the cut corresponding to an 
estimated contained event fraction of $>$ 95\%.}
  \label{fig:fov2}
\end{figure}

In addition, a set of fiducial volume cuts is applied to allow for an unbiased measurement of the $\Xmax$-distribution:
Energy dependent cuts on the zenith angle and the maximum tank-core distance ensure a single-tank trigger probability
near one for protons and iron at all energies.\\ In order to minimise systematic uncertainties from the
relative timing between the fluorescence and surface detectors, the minimum viewing angle under which
a shower was observed is required to be larger than 20$^\circ$. This cut also removes events with 
a large fraction of direct Cherenkov light.\\ 
Moreover, a minimisation of the effect of the field of view boundaries 
of the FDs is of utmost importance: The current fluorescence detectors cover an elevation range from $\Omega_1=1.5^\circ$
                 to $\Omega_2=30^\circ$ and therefore the observable heights for
                 vertical tracks are between $R\tan \Omega_1<h_v<R\tan \Omega_2$, where R denotes
                 the distance of the shower core to the fluorescence detector. That is, the farther
                 away from a fluorescence detector a track is detected, the smaller becomes the observable upper slant depth 
                 boundary $X_\mathrm{up}$. Similarly the lower slant depth boundary $X_\mathrm{low}$ becomes larger for near showers.\\
                 Since in the quality selection it is required that the $\Xmax$ is detected within the field of view, these
                 slant depth boundaries can severely bias the selected $\Xmax$-distributions, as it is sketched in Fig.~\ref{fig:fov1}. This bias can be avoided by selecting only tracks with geometries corresponding
                 to an  $X_\mathrm{up}$-$X_\mathrm{low}$ range, which is large enough to contain most of the parent $\Xmax$-distribution. 
		 Therefore, we investigate the dependence of $\meanXmax$ on the field of view boundaries and
place fiducial volume cuts at the $X_\mathrm{up}$ and $X_\mathrm{low}$ values, where the $\meanXmax$ starts to
be constant. An example of this procedure is shown in Fig.~\ref{fig:fov2}.

\section{Systematic Uncertainties}
The effect of atmospheric uncertainties on the measurement of the shower maximum
is discussed in detail in [5]. The dominating contribution is the long-term validity
of the monthly average molecular profiles used in this analysis, which we estimate
to be $\le$ 6~\gcm{}. Using a full detector and atmosphere simulation [6], the 
 profile reconstruction algorithm [7] was found to be unbiased within 5~\gcm{} 
at all energies. The effect of multiple-scattered fluorescence and Cherenkov
light was estimated to contribute about 5~\gcm{} by comparing different light collection
algorithms. \\
Re-reconstructing showers with the geometry determined from 
the surface detector data alone yields an upper bound on the 
geometrical uncertainty of $\le 6$~\gcm{}.\\ The geometrical bias due to 
the camera alignment uncertainty is below $3$~\gcm{} and the residual acceptance
difference [8] between proton and iron showers contributes around $10$~\gcm{} at lowest
energies vanishing rapidly to zero above 10$^{18}$ \eV{}.\\
The total uncertainty
is thus around $\le$~15~\gcm{} at low energies and $\le$ 11~\gcm{} above 10$^{18}$~\eV{}. Note that
in addition the current uncertainty of the FD energy scale of 22\% [3] needs to be taken into account.
\begin{figure*}
 \centering
 \includegraphics[width=.86\textwidth]{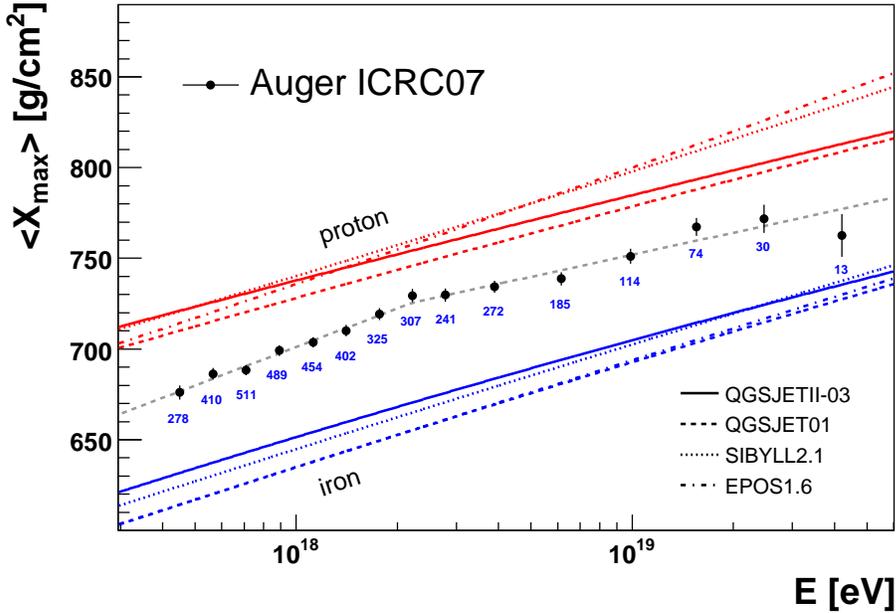}
  \caption{$\meanXmax$ as a function of energy compared to predictions from
          hadronic interaction models. The dashed line denotes a fit with two constant elongation rates and a break-point.
          Event numbers are indicated below each data point.}
  \label{fig:elong}
\end{figure*}

\section{Results}
\begin{figure}
\includegraphics[width=.49\textwidth]{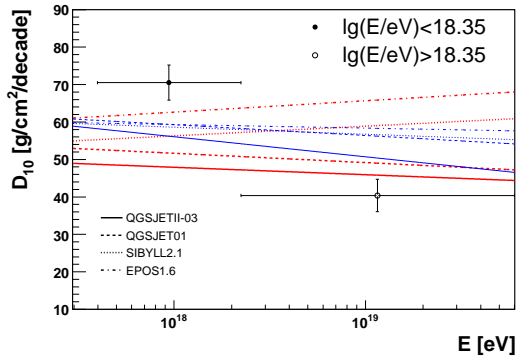}
\caption{Comparison
          of the measured elongation rate, $D_{10}$, below (solid circle) and above (open circle) 10$^{18.35}$~\eV{} to predictions of air shower
          simulations (red: protons, blue: iron).}
  \label{fig:d10}
\end{figure}

After all cuts are applied, 4329 events remain for the composition analysis. 
In Fig.~\ref{fig:elong} the
mean $\Xmax$ as a function of energy is shown along with predictions from air shower simulations [10,11]. 
As can be seen, our measurement favours a mixed composition at all energies.\\
A simple linear fit, $\meanXmax=D_{10}\cdot\lg\left(E/\!\!\eV\right)+c$, yields an elongation rate
of 54$\pm$2 (stat.)~\gcm{}/decade, but does not describe our data very well ($\chi^2/$Ndf$=24/13$, P$<$3\%).
Allowing for a break in the elongation rate at an energy $E_\mathrm{b}$ leads to a satisfactory fit with
$\chi^2/$Ndf$=9/11$, P$=$63\% and $D_{10}=71\pm5$ (stat.)~\gcm{}/decade below $E_\mathrm{b}=10^{18.35}$~\eV{} and
$D_{10}=40\pm4$ (stat.)~\gcm{}/decade above this energy. This fit is indicated as a dashed gray line
in Fig.~\ref{fig:elong}.\\
Due to the uncertainties of hadronic interaction at highest energies,
the interpretation of these elongation rates is, however, ambiguous (cf. Fig.~\ref{fig:d10}).
Using the QGSJETII elongation rates the data suggests a moderate lightening  of the
primary cosmic at low energies and an almost constant composition at high energies, 
whereas the EPOS elongation rate is clearly larger than the measured one at high energies, which
would indicate a transition from light to heavy elements. Theses ambiguities
will be partially resolved by the analysis of the $\Xmax$ fluctuations as
an additional mass sensitive parameter. \\
A comparison with previous measurements [9] is presented in Fig.~\ref{fig:comp}.
The results of all three experiments are compatible within their systematic uncertainties. It is worthwhile noting
that although the data presented here have been collected during the construction of the 
Pierre Auger Observatory, their statistical precision already exceed that of preceeding
experiments.

\section*{References}
[1]\ V.~Berezinsky et al.,
  Phys.\ Rev.\  D {\bf 74} (2006) 043005; T.~Stanev, astro-ph/0611633; A.~M.~Hillas, astro-ph/0607109, D.\  Allard et al., Astropart.Phys. 27, (2007), 61.

[2]\ J.\ Linsley, Proc.\  15th ICRC, 12 (1977) 89; 
   T.K.\  Gaisser et al., Proc.\  16th ICRC, 9 (1979) 258; 
   J.\  Linsley and A.A.\  Watson, Phys.\  Rev.\  Lett., 46 (1981) 459.

[3]\ B.\ Dawson [Pierre Auger Collaboration], these proceedings, \#0976

[4]\ T.K.~Gaisser and A.M.~Hillas, Proc. 15th ICRC (1977), 8, 353.

[5]\ M.\ Prouza [Pierre Auger Collaboration], these proceedings, \#0398

[6]\ L.\ Prado et al., Nucl.\ Instrum.\ Meth., A545 (2005), 632.
 
[7]\ M.\ Unger, R.\ Engel, F.\ Sch\"ussler, R.\ Ulrich, these proceedings, \#0972

[8]\ H.O.\ Klages  [Pierre Auger Collaboration],  these proceedings, \#0065

[9]\ D.J.\  Bird et al. [Fly's Eye Collaboration], Phys.\  Rev.\  Lett., 71 (1993) 3401; T.\  Abu-Zayyad et al. [HiRes-MIA Collaboration], Astrophys.\  J., 557 (2001) 686; R.U.\  Abbasi et al. [HiRes Collaboration], Astrophys.\  J., 622 (2005) 910.

[10]\ N.N.\  {Kalmykov et al., Nucl.\  Phys.\  B (Proc.\  Suppl.) (1997), 7; R.~Engel et al., Proc.\  26th ICRC (1999), 415; S.\  Ostapchenko, Nucl.\  Phys.\  Proc.\  Suppl., 151 (2006), 143; T.~Pierog et al., these proceedings, \#0905 and astro-ph/0611311.

[11]\ T.\ Bergmann et al., Astropart.\  Phys., 26, (2007), 420.

\begin{figure}
 \includegraphics[width=.49\textwidth]{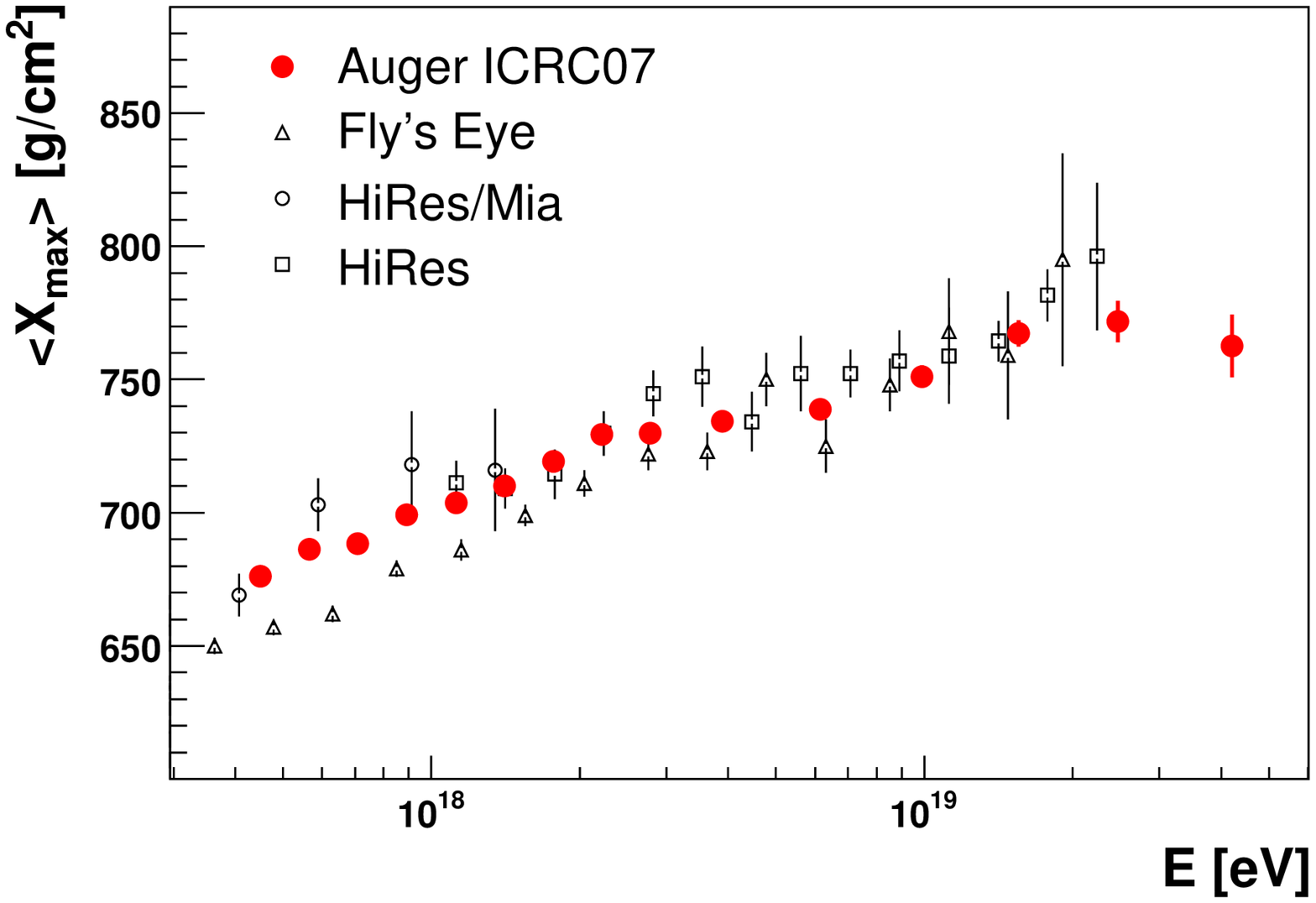}
  \caption{$\meanXmax$ as a function of energy compared to previous experiments.}
  \label{fig:comp}
\end{figure}

\end{document}